\documentclass[pdftex, twocolappendix, numberedappendix, iop, apj]{emulateapj}

\usepackage[T1]{fontenc}
\usepackage{scrextend}
\usepackage{hyperref}
\usepackage{graphicx}
\usepackage{color}
\usepackage{natbib}
\usepackage{sidecap}
\usepackage{enumitem}
\usepackage{amsmath}
\usepackage{amssymb}
\usepackage{courier}

\newcommand{\be}{\begin{equation}}
\newcommand{\ee}{\end{equation}}

\newcommand{\ba}{\begin{eqnarray}}
\newcommand{\ea}{\end{eqnarray}}

\newcommand{\COBE}{\textsl{COBE}}
\newcommand{\wmap}{\textsl{WMAP}}
\newcommand{\WMAP}{\textsl{WMAP}}
\newcommand{\planck}{\textsl{Planck}}
\newcommand{\Planck}{\textsl{Planck}}
\newcommand{\lcdm}{\ensuremath{\Lambda\mathrm{CDM}}}

\newcommand{\camb}{\texttt{CAMB}}
\newcommand{\cosmomc}{\texttt{CosmoMC}}

\shorttitle{Impact of BAO on $H_0$}
\shortauthors{G.~E.~Addison et al.} 

\begin{document}

\title{Elucidating \lcdm: Impact of Baryon Acoustic Oscillation Measurements on the Hubble Constant Discrepancy}

\author{G.~E.~Addison\altaffilmark{1}, D.~J.~Watts\altaffilmark{1}, C.~L.~Bennett\altaffilmark{1}, M.~Halpern\altaffilmark{2}, G.~Hinshaw\altaffilmark{2},  and J.~L.~Weiland\altaffilmark{1}}

\email{gaddison@jhu.edu}

\altaffiltext{1}{
Dept. of Physics \& Astronomy, The Johns Hopkins University, 3400 N. Charles St., Baltimore, MD 21218-2686
}

\altaffiltext{2}{
Department of Physics and Astronomy, University of British Columbia, 6224 Agricultural Road, Vancouver, BC V6T 1Z1, Canada
}

\begin{abstract}

We examine the impact of baryon acoustic oscillation (BAO) scale measurements on the discrepancy between the value of the Hubble constant ($H_0$) inferred from the local distance ladder and from \planck\ cosmic microwave background (CMB) data. While the BAO data alone cannot constrain $H_0$, we show that combining the latest BAO results with \wmap, Atacama Cosmology Telescope (ACT), or South Pole Telescope (SPT) CMB data produces values of $H_0$ that are $2.4-3.1\sigma$ lower than the distance ladder, independent of \planck, and that this downward pull was less apparent in some earlier analyses that used only angle-averaged BAO scale constraints rather than full anisotropic information. At the same time, the combination of BAO and CMB data also disfavors the lower values of $H_0$ preferred by the \planck\ high-multipole temperature power spectrum. Combining galaxy and Lyman-$\alpha$ forest (Ly$\alpha$) BAO with a precise estimate of the primordial deuterium abundance produces $H_0=66.98\pm1.18$~km~s$^{-1}$~Mpc$^{-1}$ for the flat \lcdm\ model. This value is completely independent of CMB anisotropy constraints and is $3.0\sigma$ lower than the latest distance ladder constraint, although $2.4\sigma$ tension also exists between the galaxy BAO and Ly$\alpha$ BAO. These results show that it is not possible to explain the $H_0$ disagreement solely with a systematic error specific to the \planck\ data. The fact that tensions remain even after the removal of any single data set makes this intriguing puzzle all the more challenging to resolve.
\end{abstract}

\keywords{
cosmic background radiation -- cosmological parameters -- cosmology: observations -- distance scale -- large-scale structure of universe
}

\section{Introduction}

While no single data set currently provides compelling evidence for a deviation from the standard Lambda cold dark matter (\lcdm) cosmological model, the values of some parameters inferred from different measurements now exhibit moderate to severe tension. This is most pronounced in the value of the Hubble constant, $H_0$. \citeauthor{riess/etal:2016} (2016; hereafter R16) provided the most recent and most precise local distance ladder constraint, finding $H_0=73.24\pm1.74$~km~s$^{-1}$~Mpc$^{-1}$ by combining three absolute distance anchors with the empirical period-luminosity relation for Cepheid variable stars and the relationship between observed light curve and intrinsic luminosity of type Ia supernovae (SNe). The most precise $H_0$ prediction from cosmic microwave background (CMB) anisotropy power spectrum measurements is currently provided by the \planck\ mission. The 2015 \planck\ temperature and polarization analysis produced $H_0=67.31\pm0.96$~km~s$^{-1}$~Mpc$^{-1}$ \citep{planck/13:2015}. An updated analysis with a revised estimate of the optical depth to reionization, $\tau$, found $H_0=66.88\pm0.91$, or $66.93\pm0.62$ if preliminary small-scale polarization data are also included \citep{planck/intermediate/46:2016}. Assuming all uncertainties are Gaussian, these values are, respectively, 3.0, 3.2, and $3.4\sigma$ lower than the distance ladder determination. Strong lensing timing delay measurements have produced $H_0$ constraints consistent with the distance ladder, and in mild tension with \planck\ \citep{bonvin/etal:2017}. Tensions also exist between the \planck\ predictions for the growth of cosmic structure (through the matter density, $\Omega_m$, and present-day density fluctuation amplitude, $\sigma_8$) and measurements using weak gravitational lensing \citep[e.g.,][]{hildebrandt/etal:2017,joudaki/etal:2017,alsing/heavens/jaffe:2017,kohlinger/etal:2017}.

It is not clear whether the problem is with the model or the data. While it is certainly plausible that a failure of the standard model could show up as a discrepancy between the CMB and low-redshift measurements, none of the commonly-considered or physically-motivated extensions to \lcdm\ appear to provide a convincing improvement when considering the full range of data available \citep[e.g.,][]{planck/13:2015,bernal/etal:2016}. In principle, the CMB prediction for $H_0$ could be significantly increased by modifying the expansion history of the universe post-recombination, for example by allowing spatial curvature or a dark energy equation of state $w\neq-1$. \planck\ temperature and polarization data alone mildly prefer a non-zero curvature, but $H_0$ goes in the wrong direction. The \planck\ 2015 \lcdm$+\Omega_k$ constraint is $53.2\pm5.1$~km~s$^{-1}$~Mpc$^{-1}$ (mean and standard deviation), with 95\% of Markov chain Monte Carlo (MCMC) samples lying in $43.7<H_0/$~km~s$^{-1}$~Mpc$^{-1}<63.5$\footnote{The public \planck\ MCMC chains can be downloaded from the \planck\ legacy archive: http://pla.esac.esa.int/pla/}. Allowing $w<-1$ can shift the \planck\ prediction to 70 or even 80~km~s$^{-1}$~Mpc$^{-1}$, however, even leaving aside questions of the physical interpretation of $w<-1$, resolving the $H_0$ disagreement with evolution in $w$ is strongly disfavored when we include observations of the expansion rate, such as baryon acoustic oscillations (BAO) in the clustering of galaxies, or high-redshift SNe. \cite{alam/etal:2017} combined \planck\ data with the latest galaxy clustering and SNe data and found $H_0=67.9\pm0.9$~km~s$^{-1}$~Mpc$^{-1}$ for constant $w$, or $67.5\pm1.0$~km~s$^{-1}$~Mpc$^{-1}$ for the $w_0-w_a$ parameterization.

Modifying the early-universe expansion history, for instance by increasing the number of effective neutrino species, $N_{\rm eff}$, can increase the CMB $H_0$ prediction. The \planck\ data do not favor this solution, for example \cite{alam/etal:2017} report $N_{\rm eff}=2.97\pm0.20$ (\planck-only), and $3.03\pm0.18$ (\planck\ plus galaxy clustering), consistent with the standard model value of $3.046$, with corresponding $H_0$ constraints of $66.6\pm1.6$ and $67.5\pm1.2$~km~s$^{-1}$~Mpc$^{-1}$. Adding $N_{\rm eff}$ in these fits shifted the tension with the distance ladder from $3.2\sigma$ to $2.8\sigma$ (\planck-only) and from $3.1\sigma$ to $2.7\sigma$ (\planck\ plus galaxy clustering). A fit to the 2015 \planck\ temperature and polarization data plus BAO fixing $N_{\rm eff}=3.4$, the value found by R16 to most effectively relieve \planck-distance ladder tension, leads to an increase in the parameter combination best constrained by weak lensing measurements, $\sigma_8\Omega_m^{0.5}$, by around 1.5\%, 0.8 times the original uncertainty\footnote{This result is taken from the \planck\ 2015 \texttt{base\_plikHM\_TTTEEE\_lowTEB\_post\_BAO} and \texttt{base\_nnu\_plikHM\_TTTEEE\_lowTEB\_nnup39\_BAO} chains.}. This slightly worsens the tension between \planck\ and the weak lensing analyses mentioned above, which found $\sigma_8\Omega_m^{0.5}$ values lower than \planck\ at the $2-3\sigma$ level when the standard model was assumed. \cite{brust/cui/sigurdson:2017} found that the \planck-lensing consistency could be improved by also introducing some degree of neutrino or dark radiation self-interaction, but, even with a second additional parameter, a joint fit to the \planck, BAO, distance ladder, weak lensing, and galaxy cluster data produced a $H_0$ distribution peaking at $69.95$~km~s$^{-1}$~Mpc$^{-1}$, still almost $2\sigma$ lower than the R16 measurement. In short, while a non-standard value of $N_{\rm eff}$ cannot be ruled out, its inclusion is not justified by the improvements to the fit.

On the other hand, the discrepant data sets have passed a range of systematic checks. The R16 distance ladder analysis used infrared data to greatly reduce the effects of reddening, substituted rungs of the ladder with alternative data, compared different calibrators, corrected for estimated local motion, and constructed a systematic error budget from considering a range of modeling variants \citep[see also, e.g.,][]{cardona/kunz/pettorino:2017,zhang/etal:2017,feeney/mortlock/dalmasso:prep,dhawan/jha/leibundgut:prep,follin/knox:prep}. The distance ladder measurements have substantially improved since the analysis by \cite{efstathiou:2014}. While the constraints have become tighter, the mean $H_0$ values in recent years have remained fairly constant \citep[e.g.,][]{riess/etal:2009,riess/etal:2011,freedman/etal:2012}. Likewise, the \planck\ team has performed an array of robustness checks of their data, investigating the effects of detector nonlinearity, beam shapes and sidelobes, and various other calibration-related issues. Also, the preference for a lower $H_0$ from \planck\ does not appear to be driven by a particular frequency channel \citep{planck/51:prep}.

Ultimately it may take additional high-precision measurements to shed light on what is really going on. More precise measurements may bring with them new tensions or disagreements, and the handling of systematic errors will get harder, not easier, as statistical uncertainties are reduced. In the meantime, it is therefore helpful to reexamine existing data and ask whether any extra insight into the discrepancies can be gleaned. To this end, in this paper we investigate in detail the indirect but important role played by BAO measurements in $H_0$ constraints, both with and without CMB anisotropy data. While this topic has been addressed in the literature, we describe several results that have either not previously been discussed, or are not widely appreciated. In Section~2, we review the BAO measurements. In Section~3, we describe results of fitting cosmological parameters to BAO in conjunction with other data sets, focussing on $H_0$. A discussion and conclusions follow in Sections~4 and 5.

\section{BAO measurements}

The first convincing detections of the BAO feature in the correlation function or power spectrum of large-scale structure (LSS) tracers were made a little over a decade ago \citep{eisenstein/etal:2005,cole/etal:2005}. Since that time, deeper surveys with orders of magnitude more galaxies, notably the Baryon Oscillation Spectroscopic Survey (BOSS\footnote{http://www.sdss3.org/surveys/boss.php}), have led to both improved precision in the BAO scale measurements over a range of redshifts, and improved analysis methodologies \citep[e.g.,][]{percival/etal:2014,anderson/etal:2014,kazin/etal:2014,reid/etal:2016}. While current and future BAO surveys are proposed as a means of improving dark energy constraints, BAO measurements also provide significant information about parameters in the standard \lcdm\ model, particularly in joint fits with the CMB.

A detailed discussion of BAO physics can be found in Chapter~4 of \cite{weinberg/etal:2013}. The BAO scale in the transverse and line-of-sight direction correspond to measurements of $D_M(z)/r_d$ and $H(z)r_d$, where $D_M(z)=(1+z)D_A(z)$ is the comoving angular diameter distance at the effective redshift of the survey and $r_d$ is the sound horizon at the drag epoch where baryons decouple from photons, denoted $z_d$. The sound horizon is defined as\footnote{The sound horizon was referred to as $r_s$ by \cite{addison/etal:2013b}, we have adopted the $r_d$ notation here for consistency with other work.}
\be
r_d=\int_{z_d}^{\infty}dz\frac{c_s(z)}{H(z)},
\ee
where the sound speed, $c_s=c/\sqrt{3(1+R)}$, depends on the ratio of baryon to photon density, with $R=3\rho_b/4\rho_{\gamma}$. The sound horizon is sensitive to the physics of the early universe, including the pre-recombination expansion history and the number of effective neutrino species, $N_{\rm eff}$, while $D_M(z)$ and $H(z)$ at the effective redshift of the survey depend on the late-time expansion.

\begin{table*}
  \centering
  \caption{BAO measurements used in this work}
  \begin{tabular}{llclcl}
\hline
Dataset&LSS tracer&$z_{\rm eff}$&Measurement\footnote{\label{first}Note that the fiducial sound horizon, $r_{d,{\rm fid.}}$, differs across different analyses. We provide constraints here only to show relative precision. For parameter fitting we use full likelihood surfaces, including correlations across the BOSS redshift bins or between $D_M$ and $H$.}&Constraint\footref{first}&Reference\\
\hline
6dFGS&galaxies&0.106&$r_d/D_V(z_{\rm eff})$&$0.336\pm0.015$&\cite{beutler/etal:2011}\\
\\
SDSS MGS&galaxies&0.15&$D_V(z_{\rm eff})\,r_{d,{\rm fid.}}/r_d$ [Mpc]&$664\pm25$&\cite{ross/etal:2015}\\
\\
BOSS DR12&galaxies&0.38&$D_M(z_{\rm eff})\,r_{d,{\rm fid.}}/r_d$ [Mpc]&$1512\pm25$&\cite{alam/etal:2017}\\
&&&$H(z_{\rm eff})\,r_d/r_{d,{\rm fid.}}$ [km~s$^{-1}$~Mpc$^{-1}$]&$81.2\pm2.4$&\\
&&0.51&$D_M(z_{\rm eff})\,r_{d,{\rm fid.}}/r_d$ [Mpc]&$1975\pm30$&\\
&&&$H(z_{\rm eff})\,r_d/r_{d,{\rm fid.}}$ [km~s$^{-1}$~Mpc$^{-1}$]&$90.9\pm2.3$&\\
&&0.61&$D_M(z_{\rm eff})\,r_{d,{\rm fid.}}/r_d$ [Mpc]&$2307\pm37$&\\
&&&$H(z_{\rm eff})\,r_d/r_{d,{\rm fid.}}$ [km~s$^{-1}$~Mpc$^{-1}$]&$99.0\pm2.5$&\\
\\
BOSS DR11 Ly$\alpha$&Ly$\alpha$ absorbers\footnote{\label{second}For brevity we refer to the Ly$\alpha$ and QSO$\times$Ly$\alpha$ measurements collectively as Ly$\alpha$.}&2.34&$D_A(z_{\rm eff})/r_d$&$11.28\pm0.65$&\cite{delubac/etal:2015}\\
&&&$c/\left[H(z_{\rm eff})r_d\right]$&$9.18\pm0.28$\\
\\
BOSS DR11 QSO$\times$Ly$\alpha$&QSO, Ly$\alpha$\footref{second}&2.36&$D_A(z_{\rm eff})/r_d$&$10.8\pm0.4$&\cite{font-ribera/etal:2014}\\
&&&$c/\left[H(z_{\rm eff})r_d\right]$&$9.0\pm0.3$&\\
\hline
\end{tabular}
\end{table*}

In some cases, only a joint constraint, for example on $D_V(z)/r_d$, where $D_V(z)=[czD_M^2(z)/H(z)]^{1/3}$, is provided, representing an angle-averaged constraint. This can be helpful where the BAO feature is detected at lower significance and the separate line-of-sight and transverse measures are poorly constrained or have distributions with non-Gaussian tails. Whenever possible, we use the joint anisotropic $D_M(z)/r_d$ and $H(z)r_d$ constraints. Quantities like $D_V(z)/r_d$ entail a compression of information that potentially give a false sense of agreement with other data, as discussed in Section~3.2.

\subsection{Current BAO constraints}

The BAO data sets included in fits presented in this paper are listed in Table~1. For the 6dF Galaxy Survey (6dFGS) and Sloan Digital Sky Survey (SDSS) Main Galaxy Sample (MGS), we adopt a simple Gaussian likelihood for $r_d/D_V(z)$ or $D_V(z)/r_d$. Away from the peak of the likelihood these constraints become non-Gaussian, however the uncertainties for these measurements are large enough that the preferred model solutions never lie far from the peak in a joint fit with other data. We use the consensus BAO scale measurements from the BOSS Data Release 12 (DR12), including $D_M(z)/r_d$ and $H(z)r_d$ for each of the three redshift bins and the six-by-six covariance matrix described by \cite{alam/etal:2017}. We restrict our analysis to the BAO scale as it is the most robust observable from LSS surveys \citep[e.g.,][and references therein]{weinberg/etal:2013}, and do not consider redshift-space distortion constraints or information from the broadband correlation function. We do not include results from the WiggleZ\footnote{http://wigglez.swin.edu.au/site/} survey, which are consistent with BOSS and partially overlap in sky coverage \citep{beutler/etal:2016}.

BAO have been measured in the Lyman-$\alpha$ (Ly$\alpha$) forest of BOSS quasars (QSOs), and in the cross-correlation between the QSOs and Ly$\alpha$ absorbers, at effective redshifts of $2.3-2.4$ \citep{busca/etal:2013,slosar/etal:2013,font-ribera/etal:2014,delubac/etal:2015,bautista/etal:2017}. BAO measurements at these redshifts, when the dark energy contribution to the total energy budget of the universe is small, are a powerful complement to the BAO from lower-redshift galaxy surveys. The analysis methodology and systematic error treatment required to extract the Ly$\alpha$ BAO scale are less mature than for the galaxy BAO and are an active field of research \citep[e.g.,][]{blomqvist/etal:2015}. The anisotropic BAO measurements from the DR11 Ly$\alpha$ and QSO$\times$Ly$\alpha$ analyses are in $\sim2.5\sigma$ tension with \planck\ predictions assuming a standard flat \lcdm\ model. This tension was reduced slightly in the DR12 Ly$\alpha$ BAO analysis \citep{bautista/etal:2017}. \cite{bautista/etal:2017} found that the shift in the DR12 Ly$\alpha$ constraints was predominantly due to the additional data rather than some different treatment of systematic effects\footnote{The DR12 QSO$\times$Ly$\alpha$ analysis, released while this work was in review, produced results consistent with DR11, in tension with \planck\ predictions at the $2.3\sigma$ level \citep{bourboux/etal:prep}}. We present results using the DR11 Ly$\alpha$ and QSO$\times$Ly$\alpha$ constraints, and from combining the galaxy and Ly$\alpha$ BAO, noting that $\sim2.5\sigma$ effects can and do arise purely from statistical fluctuations, and that there is currently no known systematic error that explains this tension.

Other BAO measurements have been reported, for example using galaxy clusters as LSS tracers \citep[e.g.,][]{veropalumbo/etal:2016,hong/etal:2016}. These results are generally less precise than the galaxy BAO, at similar redshifts, and their inclusion would not significantly affect our results. Recently, the first measurement of BAO from the extended Baryon Oscillation Spectroscopic Survey (eBOSS\footnote{http://www.sdss.org/surveys/eboss/}) was reported using clustering of quasars at $0.8\leq z\leq2.2$ \citep{ata/etal:prep}. BAO constraints from this redshift range are potentially a useful addition to the galaxy and Ly$\alpha$ BAO and upcoming, higher-precision eBOSS measurements will be interesting to include in future analyses.

\subsection{Choice of CMB data for joint fits}

Joint fits between \planck\ and BAO have been reported extensively for a range of cosmological models  in recent work \citep[e.g.,][]{aubourg/etal:2015,planck/13:2015,alam/etal:2017}. While \planck\ provides the most precise CMB constraints, $\sim2.5\sigma$ tension exists between determination of some parameters from splitting the \planck\ power spectrum into multipoles $\ell<800$ and $\ell>800$, where the choice of 800 corresponds to a roughly even division of overall constraining power \citep{addison/etal:2016}. In the full \lcdm\ model space, the tension is not significant ($1.8\sigma$ for the assumptions used by \citeauthor{addison/etal:2016} 2016; see also \citeauthor{planck/51:prep} 2016). Current low-redshift cosmological observations do not provide strong constraints across the full \lcdm\ parameter space, however they do provide independent and precise constraints on a subset of parameters, including $H_0$, $\Omega_m$, and $\sigma_8$. These parameters are therefore of particular interest when it comes to assessing the performance of the \lcdm\ model and testing for alternatives. Given the moderate internal \planck\ tension in these parameters, it is informative to consider other CMB measurements to help understand the extent to which conclusions are driven by \planck\ data, or are independent of \planck. In this work we therefore also include results from the final \wmap\ 9-year analysis \citep{bennett/etal:2013,hinshaw/etal:2013}, the Atacama Cosmology Telescope polarization-sensitive receiver \citep[ACTPol;][]{thornton/etal:2016,louis/etal:2017,sherwin/etal:prep} two-season survey, covering 548~deg$^2$, and the 2500~deg$^2$ South Pole Telescope Sunyaev-Zel'dovich survey \citep[SPT-SZ;][]{carlstrom/etal:2011,story/etal:2013,vanengelen/etal:2012}.

\begin{table*}
  \centering
  \caption{Constraints on $H_0$ in the \lcdm\ model from the CMB alone and from combining CMB with BAO data, with the significance of the difference from the distance ladder measurement \citep[$73.24\pm1.74$;][]{riess/etal:2016} in parenthesis, assuming uncorrelated Gaussian errors (all values in km~s$^{-1}$~Mpc$^{-1}$)}
  \begin{tabular}{lllccc}
\hline
CMB dataset&Large-scale likelihood\footnote{Pixel-based and other likelihoods used at multipoles $\ell\lesssim30$. For some fits, particularly with the ACTPol and SPT experiments that do not probe these scales, we indicate the Gaussian prior adopted on $\tau$ instead.}&Power spectrum likelihoods\footnote{Temperature, E-mode polarization, temperature-polarization cross-spectrum and lensing potential power spectra are denoted TT, EE, TE, and $\phi\phi$, respectively.}&$H_0$ (CMB only)&BAO data\footnote{`gal' refers to galaxy BAO; `Ly$\alpha$' refers to Lyman-$\alpha$ forest and QSO$\times$Ly$\alpha$ BAO (see Table~1).}&$H_0$ (CMB+BAO)\\
\hline
\WMAP\ 9-year&\wmap&TT, TE, EE&$69.68\pm2.17$ $(1.3\sigma)$&gal+Ly$\alpha$&$68.30\pm0.72$ $(2.6\sigma)$\\
&\textquotedbl&\textquotedbl&\textquotedbl&gal&$68.19\pm0.72$ $(2.7\sigma)$\\
&\textquotedbl&\textquotedbl&\textquotedbl&Ly$\alpha$&$71.01\pm2.10$ $(0.8\sigma)$\\
\\
ACTPol Two-Season&$\tau=0.07\pm0.02$&TT, TE, EE, $\phi\phi$&$67.12\pm2.67$ $(1.9\sigma)$&gal+Ly$\alpha$&$67.23\pm0.80$ $(3.1\sigma)$\\
&\textquotedbl&\textquotedbl&\textquotedbl&gal&$66.94\pm0.77$ $(3.3\sigma)$\\
&\textquotedbl&\textquotedbl&\textquotedbl&Ly$\alpha$&$69.59\pm2.61$ $(1.3\sigma)$\\
&\textquotedbl&TT, TE, EE&$67.60\pm3.56$ $(1.4\sigma)$&gal+Ly$\alpha$&$67.29\pm0.83$ $(3.1\sigma)$\\
&$\tau=0.055\pm0.009$&TT, TE, EE, $\phi\phi$&$66.55\pm2.59$ $(2.1\sigma)$&gal+Ly$\alpha$&$67.21\pm0.83$ $(3.1\sigma)$\\
\\
SPT-SZ&$\tau=0.07\pm0.02$&TT, $\phi\phi$&$71.38\pm3.09$ $(0.5\sigma)$&gal+Ly$\alpha$&$68.52\pm0.90$ $(2.4\sigma)$\\
&\textquotedbl&\textquotedbl&\textquotedbl&gal&$68.25\pm0.91$ $(2.5\sigma)$\\
&\textquotedbl&\textquotedbl&\textquotedbl&Ly$\alpha$&$73.74\pm2.84$ $(0.2\sigma)$\\
&\textquotedbl&TT&$73.20\pm3.54$ $(0.0\sigma)$&gal+Ly$\alpha$&$68.49\pm0.92$ $(2.4\sigma)$\\
&$\tau=0.055\pm0.009$&TT, $\phi\phi$&$70.67\pm3.06$ $(0.7\sigma)$&gal+Ly$\alpha$&$68.46\pm0.88$ $(2.5\sigma$)\\
\\
\Planck&\texttt{lowTEB}\footnote{\texttt{lowTEB} is the combined temperature-plus-polarization \planck\ likelihood for $\ell<30$.}&\texttt{plikHM\_TT} 2015, $\phi\phi$&$67.86\pm0.92$ $(2.7\sigma)$&gal+Ly$\alpha$&$68.06\pm0.56$ $(2.8\sigma)$\\
&\textquotedbl&\textquotedbl&\textquotedbl&gal&$67.95\pm0.54$ $(2.9\sigma)$\\
&\textquotedbl&\textquotedbl&\textquotedbl&Ly$\alpha$&$68.17\pm0.93$ $(2.6\sigma)$\\
&\texttt{lowTEB}&\texttt{plikHM\_TT} 2015&$67.81\pm0.92$ $(2.8\sigma)$&gal+Ly$\alpha$&$67.97\pm0.56$ $(2.9\sigma)$\\
&$\tau=0.055\pm0.009$, \texttt{lowl}\footnote{Since the \cite{planck/51:prep} low-multipole polarization likelihood is not publicly available we approximate its inclusion with a prior $\tau=0.055\pm0.009$, which produces constraints in very good agreement with their Table~8. \texttt{lowl} is the \planck\ temperature-only likelihood for $\ell<30$ (no polarization).}&\texttt{plikHM\_TT} 2015&$66.88\pm0.91$ ($3.2\sigma$)&gal+Ly$\alpha$&$67.72\pm0.54$ $(3.0\sigma)$\\
&$\tau=0.055\pm0.009$, \texttt{lowl}&\texttt{plikHM\_TTTEEE} 2015&$66.93\pm0.62$ ($3.4\sigma$)&gal+Ly$\alpha$&$67.53\pm0.45$ $(3.2\sigma)$\\
\\
&$\tau=0.07\pm0.02$, \texttt{lowl}&\texttt{plikHM\_TT} $\ell<800$&$70.08\pm1.96$ $(1.2\sigma)$&gal+Ly$\alpha$&$68.34\pm0.67$ $(2.6\sigma)$\\
&$\tau=0.055\pm0.009$, \texttt{lowl}&\texttt{plikHM\_TT} $\ell<800$&$69.78\pm1.86$ $(1.4\sigma)$&gal+Ly$\alpha$&$68.29\pm0.66$ $(2.7\sigma)$\\
&$\tau=0.07\pm0.02$&\texttt{plikHM\_TT} $\ell>800$&$65.12\pm1.45$ $(3.6\sigma)$&gal+Ly$\alpha$&$67.91\pm0.66$ $(2.9\sigma)$\\
&$\tau=0.055\pm0.009$&\texttt{plikHM\_TT} $\ell>800$&$64.30\pm1.31$ $(4.1\sigma)$&gal+Ly$\alpha$&$67.55\pm0.62$ $(3.1\sigma)$\\
\hline
\end{tabular}
\end{table*}

\section{Results}

\subsection{Combining BAO and CMB anisotropy measurements}

In Table~2 we show $H_0$ constraints within the \lcdm\ model from CMB data sets with and without the inclusion of the BAO data. ACTPol and SPT use \wmap\ or \planck\ data only to provide an absolute calibration, that is, a single scale-independent multiplicative rescaling of the measured power spectrum. Since these experiments do not measure $\tau$, we adopt a Gaussian prior, either the same broader $\tau=0.07\pm0.02$ prior used by \cite{planck/11:2015} and \cite{addison/etal:2016}, or the $\tau=0.055\pm0.009$ constraint from the latest \planck\ HFI low-$\ell$ polarization determination \citep{planck/intermediate/46:2016}. Here and throughout this paper we quote the mean and standard deviation from MCMC runs using the \cosmomc\footnote{http://cosmologist.info/cosmomc/} package \citep{lewis/bridle:2002}, with convergence criterion $R-1<0.01$ \citep{gelman/rubin:1992}. Since we are not investigating foreground modeling in this work we use foreground-marginalized CMB likelihood codes for ACTPol and SPT \citep{dunkley/etal:2013,calabrese/etal:2013}. Uncertainties in foreground and other nuisance parameters propagate to cosmological parameters through an increase in power spectrum uncertainties in these codes. In the \planck\ rows of Table~2 we include the exact name of the likelihood file for clarity since a range of likelihoods have been provided by the \planck\ collaboration. These likelihoods include \planck\ foreground and nuisance parameters as described by \cite{planck/11:2015}. We show results with and without CMB lensing power spectra (denoted `$\phi\phi$' in the third column of Table~2), noting that the lensing measurements have a moderate effect on some of the CMB-only constraints but reduced impact when the BAO are included. In the last four rows of Table~2 we also list constraints from splitting the \planck\ temperature power spectrum at $\ell=800$ \citep{addison/etal:2016}, as mentioned in Section~2.2 and discussed further in Section~4.

Adding galaxy BAO to any of the CMB measurements listed in Table~2 substantially tightens the $H_0$ prediction, by more than a factor of three in the case of ACTPol or SPT. While there is still scatter in the CMB + galaxy BAO $H_0$ values, the spread is substantially reduced compared to the CMB-only column. The ACTPol+BAO and SPT+BAO combinations produce $H_0$ constraints of comparable precision to \planck\ alone. The synergy between the galaxy BAO and CMB measurements for \lcdm\ is illustrated in Figure~1 using the BOSS DR12 anisotropic BAO measurements at $z_{\rm eff}=0.61$ as an example. The predictions from the CMB are shown with MCMC samples color-coded by $H_0$, which varies fairly monotonically along the degeneracy line set by the angular acoustic scale, corresponding to the peak spacing in the CMB power spectrum. The MCMC samples shown are drawn from the full chains, and include points from the tails of the distributions in addition to high-likelihood samples. The shaded blue contours correspond to the BOSS measurements, which are precise enough to substantially reduce the range of $H_0$ values allowed by breaking CMB degeneracies. The mixing of colors visible in the ACTPol and SPT panels reflects additional degeneracy between $H_0$ and other parameters arising from the more limited range of angular scales provided by these data.

\begin{figure*}
\centering
\includegraphics{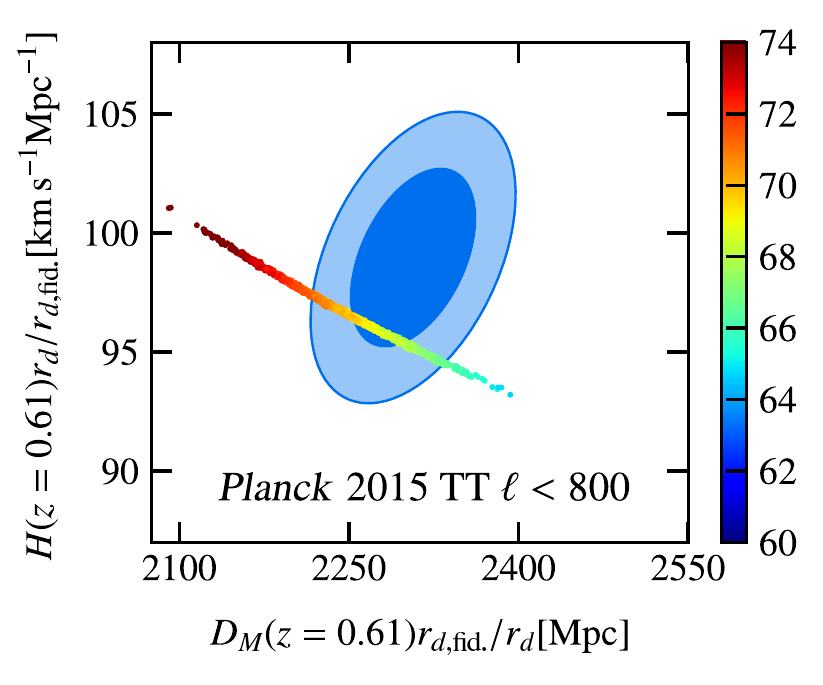}
\includegraphics{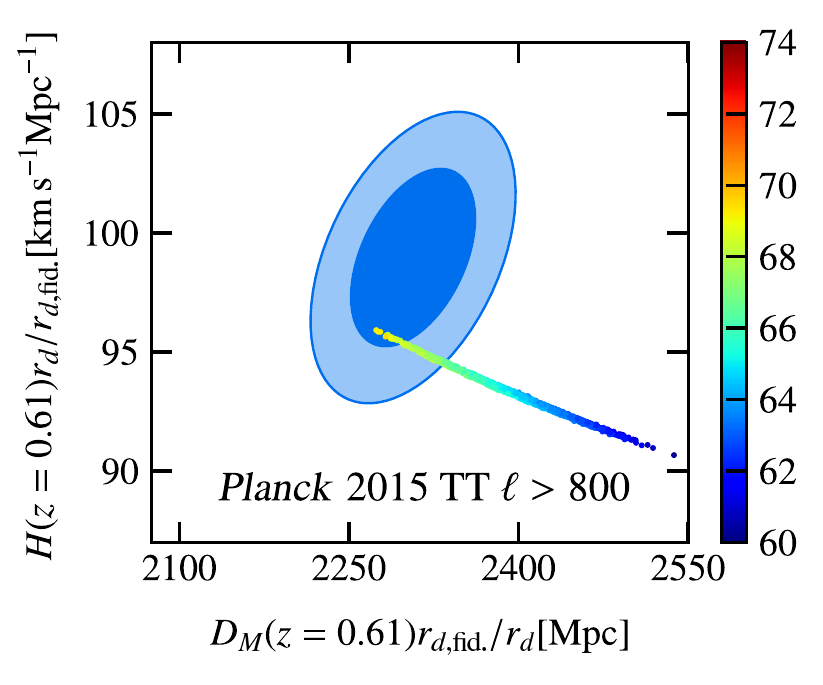}\\
\includegraphics{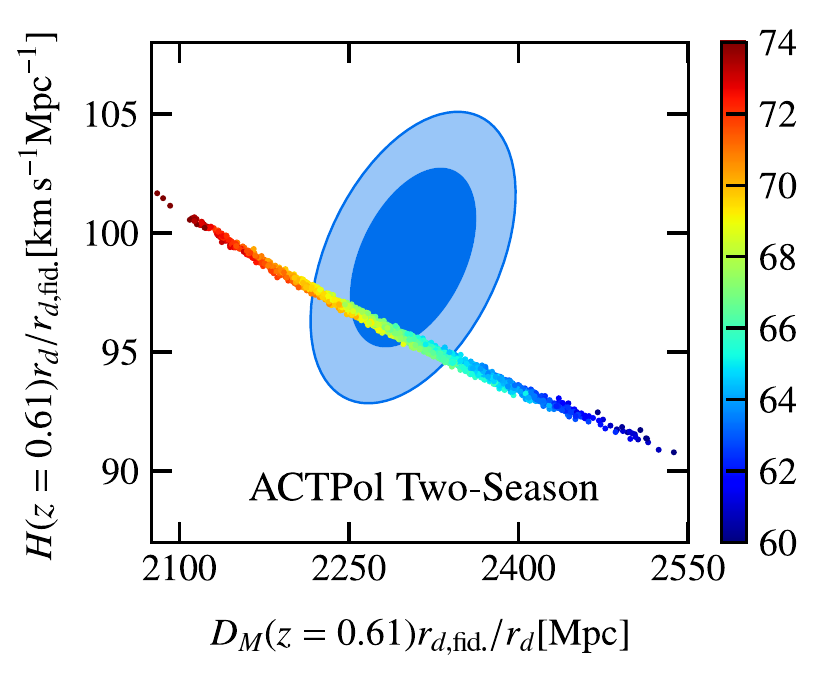}
\includegraphics{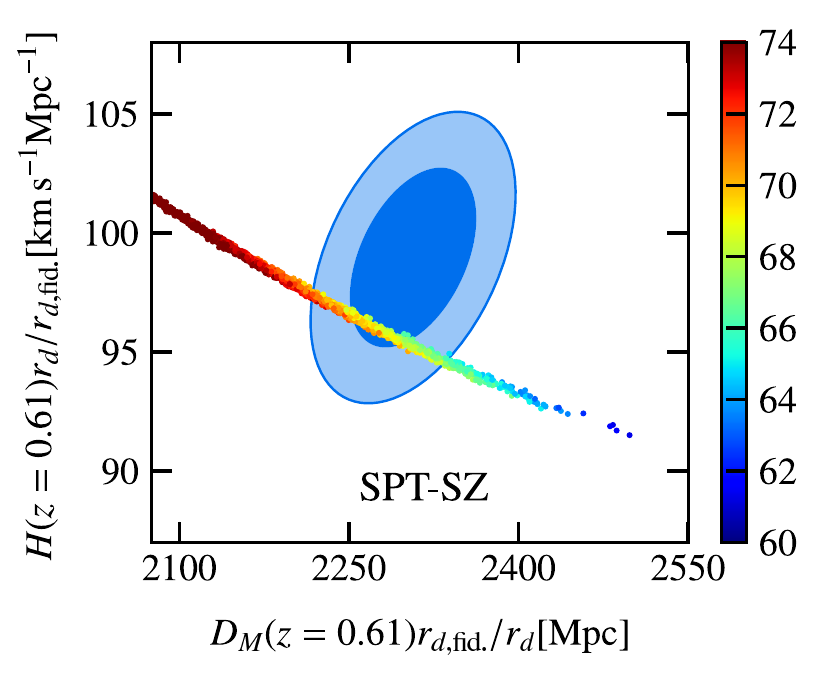}
\caption{Including BAO data substantially tightens CMB constraints on $H_0$. The observables corresponding to the transverse and line-of-sight BAO scale, $D_M\,r_{d, \textrm{fid.}}/r_d$, and $H\,r_d/r_{d,\textrm{fid.}}$ (Section~2 and Table~1), are shown for redshift $z=0.61$. The blue shaded contours are the measurements from the final BOSS DR12 analysis \citep{alam/etal:2017}. The different panels contain predictions from different, essentially independent, CMB measurements assuming a flat \lcdm\ model, with MCMC samples color-coded by $H_0$ in km~s$^{-1}$~Mpc$^{-1}$. The same $\tau=0.07\pm0.02$ prior is used in each case. The addition of the BAO tightens the $H_0$ constraint by more than a factor of three in the case of ACTPol or SPT data (Table~2). When combined with {\em any} current CMB data set the galaxy BAO disfavor the values of $H_0$ preferred by the distance ladder \citep[$73.24\pm1.74$~km~s$^{-1}$~Mpc$^{-1}$;][]{riess/etal:2016} at moderate to high significance. The lower values preferred by the high-multipole \planck\ data (the constraint from the samples shown in the top-right panel is $65.12\pm1.45$~km~s$^{-1}$~Mpc$^{-1}$) are also disfavored.}
\end{figure*}

In conjunction with the CMB, and in the context of \lcdm, the BAO have the effect of disfavoring both the higher values of $H_0$ preferred by the distance ladder, and the lower values preferred by the \planck\ damping tail at $\ell>800$. If we exclude \planck, the CMB + BAO values lie $2.4-3.1\sigma$ from the R16 measurement, depending on the choice of CMB dataset. While this trend has been reported before using \wmap\ data \citep{planck/13:2015,bernal/etal:2016}, here we show that the measurements of the damping tail from ACTPol and SPT produce the same effect even without information from the larger scales measured by the satellite experiments. The fact that combining ACTPol and BAO data produces an $H_0$ value $>3\sigma$ lower than R16 provides strong evidence that the $H_0$ discrepancy cannot be explained solely by a systematic specific to the \planck\ data. On the other hand, using the difference-of-covariance method described in Section~4.1 of \cite{planck/13:2015}, the shift in $H_0$ from adding the BAO to the $\ell>800$ \planck\ temperature power spectrum is larger than expected at the 2.2 and $2.8\sigma$ level for the $\tau=0.07\pm0.02$ and $0.055\pm0.009$ priors, respectively.

The CMB + Ly$\alpha$ BAO fits yield higher values of $H_0$ than the CMB alone, without significantly smaller uncertainties. This reflects the tension between the CMB and Ly$\alpha$ BAO discussed in earlier work \citep[e.g.,][]{delubac/etal:2015,planck/13:2015}. In a joint fit with all the BAO data the Ly$\alpha$ measurements lack the constraining power to overcome the galaxy BAO, and consequently our results are fairly insensitive to whether the Ly$\alpha$ are included along with the galaxy BAO or not. The interaction between the galaxy and Ly$\alpha$ BAO constraints is discussed further in Section~3.3.

We note that the SPT values in Table~2 differ from the value of $75.0\pm3.5$~km~s$^{-1}$~Mpc$^{-1}$ quoted in Table~3 of the original SPT analysis by \cite{story/etal:2013}. This difference is driven by three effects: (i) the inclusion of the SPT lensing $\phi\phi$ power spectrum measurement from \cite{vanengelen/etal:2012} in some of our fits, (ii) the difference in $\tau$ prior: we used $0.07\pm0.02$ or $0.055\pm0.009$, while \cite{story/etal:2013} used $0.088\pm0.015$, and (iii) different \cosmomc\ versions or fitting options, including the fact that we set the total neutrino mass to 0.06~eV in our fits, while \cite{story/etal:2013} assumed massless neutrinos, which leads to a $\sim0.2\sigma$ shift in $H_0$. We have verified that if we use the \cite{story/etal:2013} assumptions we reproduce their $75.0\pm3.5$ constraint. \cite{aylor/etal:prep} recently derived parameters from SPT with an updated \planck-based calibration and improved likelihood, however the shift they report in $H_0$ is small and would not meaningfully affect our results.

\subsection{Angle-averaged versus anisotropic BAO}

\citeauthor{bennett/etal:2014} (2014; hereafter B14) used pre-\planck\ CMB data along with BAO measurements available at the time (6dFGS, BOSS DR11, including the Ly$\alpha$ but not QSO$\times$Ly$\alpha$ cross-correlation; we refer to these data as BAO14) to constrain
\be
\begin{split}
H_0=69.3\pm0.7~\textrm{km}~\textrm{s}^{-1}~\textrm{Mpc}^{-1}\\
\textrm{ (\WMAP+ACT+SPT+BAO14).}
\end{split}
\ee
This value is noticeably higher than the CMB+BAO values reported in Table~2. To make a more direct comparison we performed an updated fit using \wmap, ACTPol, SPT, and the latest BAO data, and find
\be
\begin{split}
H_0=68.34\pm0.61~\textrm{km}~\textrm{s}^{-1}~\textrm{Mpc}^{-1}\\
\textrm{ (\WMAP+ACTPol+SPT+BAO).}
\end{split}
\ee
The difference in these values appears large given the overlap in data sets used and so we investigated this difference in detail. We found that the downward shift in our current fits is due to a combination of several effects:
\begin{enumerate}[label=(\roman*)]
\item The biggest difference comes from using the transverse and line-of-sight BOSS BAO scale measurements now available separately rather than the angle-averaged $D_V(z)/r_d$ used in B14. Using the BOSS DR11 CMASS anisotropic BAO instead of the BOSS DR11 CMASS angle-averaged BAO shifts the \wmap9+ACT+SPT+BAO14 $H_0$ constraint downwards by 0.61~km~s$^{-1}$~Mpc$^{-1}$, a shift comparable to the total uncertainty. This is discussed in more detail below.
\item A smaller shift of around 0.2~km~s$^{-1}$~Mpc$^{-1}$ is due to different likelihood codes. We find $H_0=69.07\pm0.70$~km~s$^{-1}$~Mpc$^{-1}$ using \wmap9+ACT+SPT+BAO14. Our results were obtained with the November 2016 versions of \camb\footnote{http://camb.info/} and \cosmomc, while a different MCMC code was used in B14. Furthermore, our implementation of the DR11 Ly$\alpha$ BAO constraint uses the $\chi^2$ look-up tables provided by BOSS\footnote{http://darkmatter.ps.uci.edu/baofit/}, whereas B14 constructed a likelihood directly from values reported by \cite{delubac/etal:2015}.
\item The ACTPol data have a stronger downward pull on $H_0$ than ACT. Both ACT and ACTPol prefer a lower $H_0$ value than \wmap\ alone \citep{sievers/etal:2013,louis/etal:2017}. The SPT data prefer a higher $H_0$ value than \wmap, and this preference wins out in the combination with ACT. With ACTPol, however, the downward pull is stronger, and the resulting constraint shifts downwards from $69.98\pm1.58$ (\wmap9+ACT+SPT) to $69.08\pm1.37$~km~s$^{-1}$~Mpc$^{-1}$ (\wmap+ACTPol+SPT). In combination with the BAO the impact of using ACTPol instead of ACT is subdominant to the choice of BAO constraints.

\begin{figure*}
\centering
\includegraphics{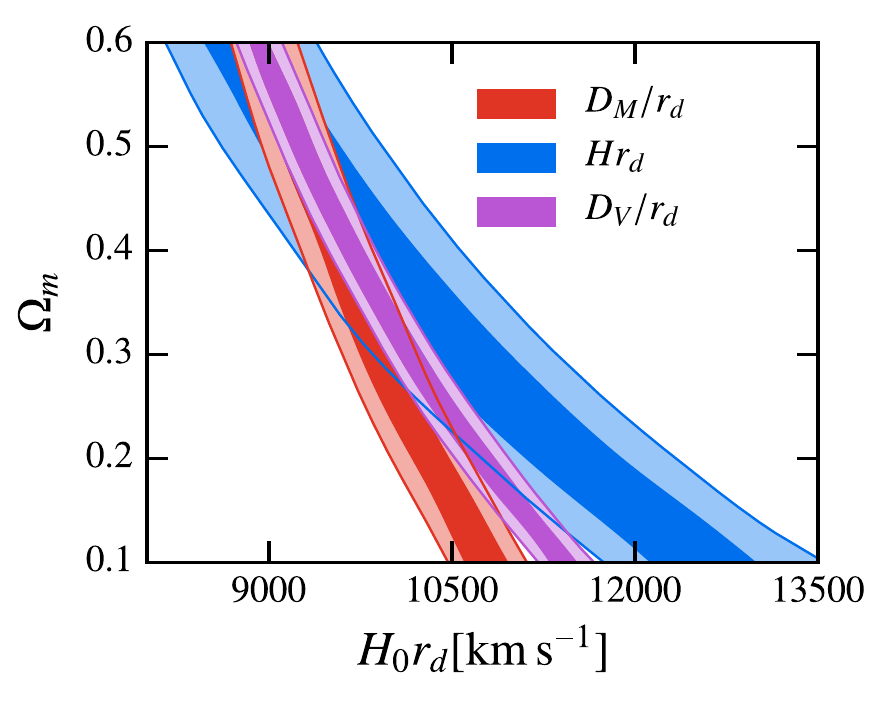}\includegraphics{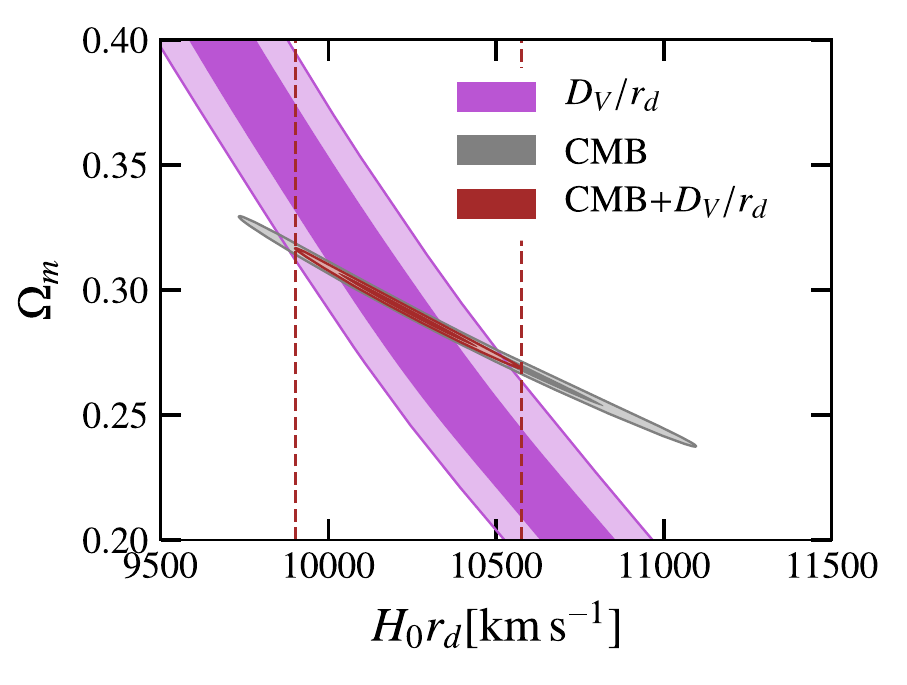}\\
\includegraphics{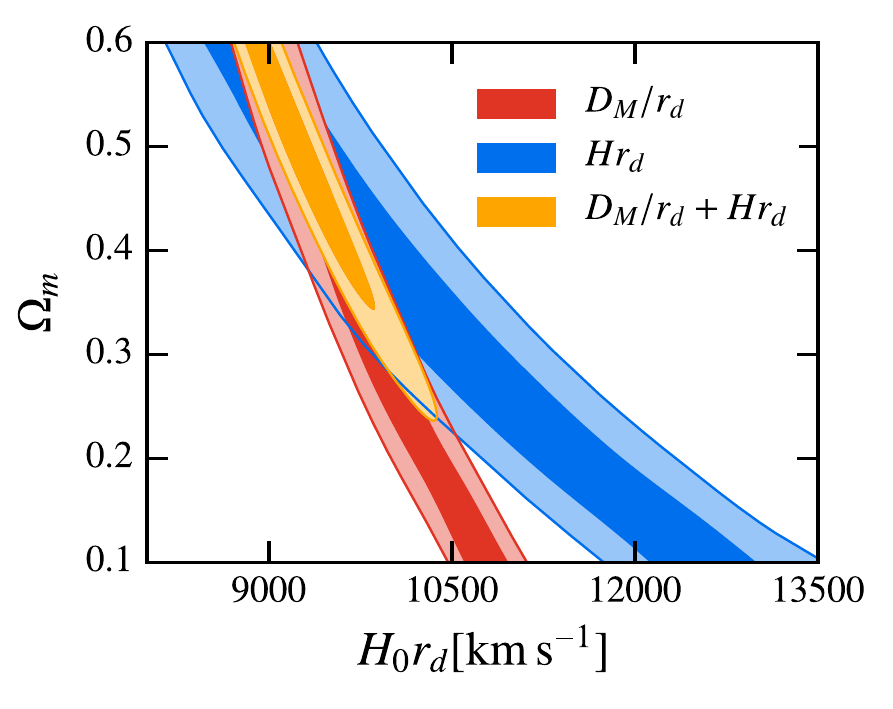}\includegraphics{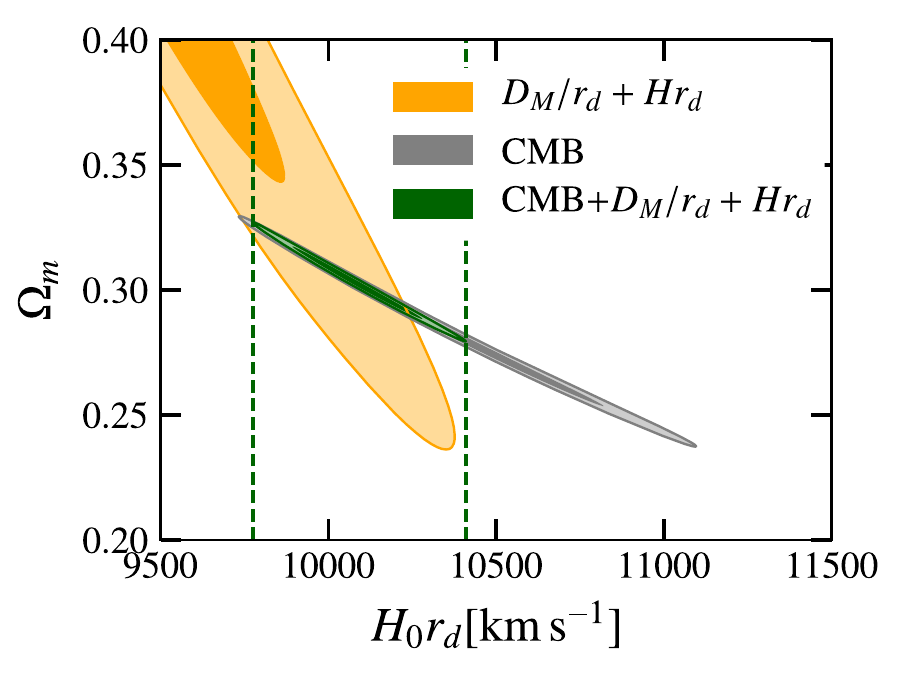}
\caption{Use of the angle-averaged BAO constraint, $D_V(z)/r_d$, instead of the full anisotropic information, $D_M(z)/r_d$ plus $H(z)r_d$, can impact determination of $H_0$ from combined CMB+BAO fits. The upper left panel shows constraints from the same BOSS CMASS DR11 galaxy sample at $z_{\rm eff}=0.57$ \citep{anderson/etal:2014} but different BAO measures -- transverse ($D_M/r_d$), line-of-sight ($Hr_d$), and angle-averaged ($D_V/r_d$, see Section~2.1). The lower left panel shows the anisotropic constraint from combining $D_M(z)/r_d$ and $H(z)r_d$. While there is significant overlap between the angle-averaged and anisotropic contours, the angle-averaged contour extends to lower values of $\Omega_m$, which are not allowed by the anisotropic constraint. The upper right and lower right panels show the effect of adding the BAO information to CMB data \citep[we show the same data sets used by][\wmap+ACT+SPT]{bennett/etal:2014}. The use of the angle-averaged $D_V(z)/r_d$ constraint diminishes the downward pull on $H_0r_d$, and also $H_0$, from the BAO. The vertical dashed lines correspond to the bounds of the contours containing 95\% of the CMB+BAO MCMC samples.}
\end{figure*}

\item The SDSS MGS BAO constraint at $z_{\rm eff}=0.15$ was not used by B14. While the MGS measurement has lower precision than BOSS (4\% compared to around 1\%), it also has a stronger preference for lower $H_0$ in conjunction with the CMB data. 
\end{enumerate}

Why does the choice of anisotropic or angle-averaged BOSS CMASS BAO make such a large difference given the same galaxy sample is used for each? In the flat \lcdm\ model, all the information from any BAO measurement is contained in contours in the two-dimensional $\Omega_m-H_0r_d$ space \citep{addison/etal:2013b}. The relative late-time expansion history is determined by $\Omega_m$, with $\Omega_{\Lambda}$ determined implicitly by the flatness constraint. The impact of radiation on the late-time expansion is small enough compared to the precision of current BAO measurements that uncertainties in the CMB temperature, which constrains the physical density $\Omega_rh^2$, or in converting to the fractional density, $\Omega_r$, can be neglected. The combination $H_0r_d$ provides an overall normalization factor and reflects the fact that the absolute length of the sound horizon, and a change in the normalization of the expansion rate, are completely degenerate when only fitting to measurements of the BAO scale.

The upper left and lower left panels of Figure~2 shows constraints in the $\Omega_m-H_0r_d$ plane for the DR11 BOSS CMASS sample at $z_{\rm eff}=0.57$ \citep{anderson/etal:2014}. We show the transverse  ($D_M/r_d$) and line-of-sight ($Hr_d$) contours separately, as well as the contour from combining both, and the angle-averaged $D_V(z)/r_d$ contour. While there is substantial overlap between the combined anisotropic contour and the $D_V(z)/r_d$ contour, a portion of the parameter space is allowed by $D_V(z)/r_d$ but ruled out by the combined anisotropic measurements. This portion is relevant when the BAO and CMB are combined, as shown in the upper right and lower right panels of Figure~2, with the anisotropic $D_M(z)/r_d+H(z)r_d$ constraint pulling down more strongly on $H_0r_d$, and hence $H_0$, since $H_0$ and $r_d$ are only partially degenerate in the CMB. The same effect is apparent in Figure~8 of \cite{cuesta/etal:2016}.

We conclude that the shift in $H_0$ from using the angle-averaged $D_V(z)/r_d$ instead of the full anisotropic BAO information is not caused by an inconsistency in the BAO measurements, but instead due to the compression of information inherent to $D_V(z)/r_d$. It is therefore preferable to use the anisotropic constraints where possible.

\subsection{Constraints from the BAO scale alone}

We now consider constraints from the BAO data without the strong additional constraining power of the CMB anisotropy measurements. As discussed above, in the flat \lcdm\ model, BAO measurements provide contours in the $\Omega_m-H_0r_d$ plane. Combining the galaxy and Ly$\alpha$ BAO provides a tight constraint on $\Omega_m$ from the late-time expansion history, even when marginalizing over the normalization $H_0r_d$. For the BAO listed in Table~1 we find constraints of
\be
\begin{split}
\Omega_m&=0.292\pm0.020\\
H_0r_d&=(10119\pm138)\textrm{~km~s}^{-1}.
\end{split}
\ee

The left panel of Figure~3 shows constraints from the galaxy and Ly$\alpha$ BAO in the $\Omega_m-H_0r_d$ plane. The orientation of these contours can be approximately understood from considering the redshift dependence of $H(z)$. Similar arguments hold for $D_A(z)$. At the Ly$\alpha$ BAO redshifts the universe is matter dominated, and $H(z)\simeq H_0\Omega_m^{1/2}(1+z)^{3/2}$, so that $H(z)r_d$ constraints produce contours along the direction with $H_0r_d\cdot\Omega_m^{1/2}$ roughly constant. At lower redshifts, where dark energy becomes dominant, $H(z)$ depends less strongly on $\Omega_m$, leading to the galaxy BAO contour being oriented more along the direction of the y-axis in Figure~3\footnote{If BAO measurements at $z=0$ were possible they would produce exactly vertical contours in Figure~3.}. There is little overlap between the galaxy and Ly$\alpha$ contours. To quantify this difference, we consider the test described in Section~4.1 of \cite{hou/etal:2014}. We calculate $\Delta\chi^2=\chi^2_{X+Y}-\chi^2_X-\chi^2_Y$, where in this case $X$ and $Y$ are the galaxy and Ly$\alpha$ BAO data, respectively,  $\chi^2_{X+Y}$ denotes the best-fit $\chi^2$ from the joint fit, and $\chi^2_X$ and $\chi^2_Y$ are the best-fit $\chi^2$ from the fits to just the galaxy or just the Ly$\alpha$ data. For Gaussian-distributed data\footnote{This is a reasonable approximation when the Ly$\alpha$ and QSO$\times$Ly$\alpha$ BAO are combined \citep{delubac/etal:2015}.}, if $X$ and $Y$ are independent and \lcdm\ is the correct model then $\Delta\chi^2$ is drawn from a $\chi^2$ distribution with $N_{\Delta\chi^2}=N_{X+Y}-N_X-N_Y$ degrees of freedom (dof). We find
\begin{align*}
\chi^2_{\rm gal}&=2.98 & N_{\rm gal}&=8-2=6\\
\chi^2_{\rm Ly\alpha}&=0.92 & N_{\rm Ly\alpha}&=4-2=2\\
\chi^2_{\rm gal+Ly\alpha}&=13.63 & N_{\rm gal+Ly\alpha}&=12-2=10\\
\Delta\chi^2&=9.73 & N_{\Delta\chi^2}&=10-6-2=2
\end{align*}
The probability to exceed (PTE) for $\chi^2=9.73$ with $N_{\rm dof}=2$ is $7.71\times10^{-3}$, which corresponds to a $2.4\sigma$ disagreement. This is comparable to the $2.5\sigma$ tension reported between the Ly$\alpha$ BAO and \planck\ measurements by \cite{delubac/etal:2015}. As discussed by \cite{aubourg/etal:2015}, modifying the cosmological model to improve three-way agreement between CMB, galaxy BAO, and Ly$\alpha$ BAO appears difficult. Here we note that the combined contour in Figure~3 lies at the intersection of the main degeneracy directions determined by the redshift coverage of the galaxy and Ly$\alpha$ measurements. If future data shift the galaxy or Ly$\alpha$ BAO constraints {\em along} these degeneracy lines (as opposed to perpendicular to them) the main result would be to change the quality of the combined fit rather than changing the parameter values. We further note that the matter density reported in (4) is in agreement with the value of $0.295\pm0.034$ from a joint analysis of type Ia SNe from several surveys covering $0<z<1$, completely independent of LSS clustering \citep{betoule/etal:2014}. This is illustrated in the right panel of Figure~3, which shows a comparison of BAO, \wmap\ 9-year, \planck\ 2016\footnote{We refer to the combination of the 2015 TT constraints with updated $\tau=0.055\pm0.009$ prior from \cite{planck/51:prep} as `\planck\ 2016'.}, and SNe constraints on $\Omega_m$ for the flat \lcdm\ model.

\begin{figure*}
\hspace{-0.2cm}
\includegraphics{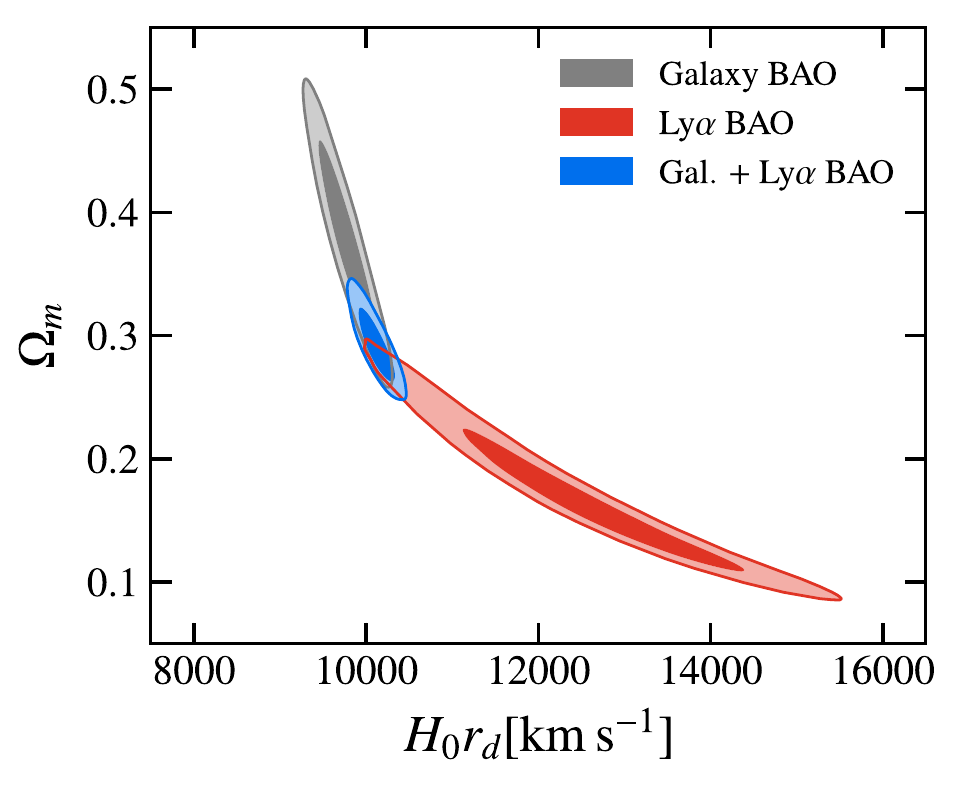}
\includegraphics{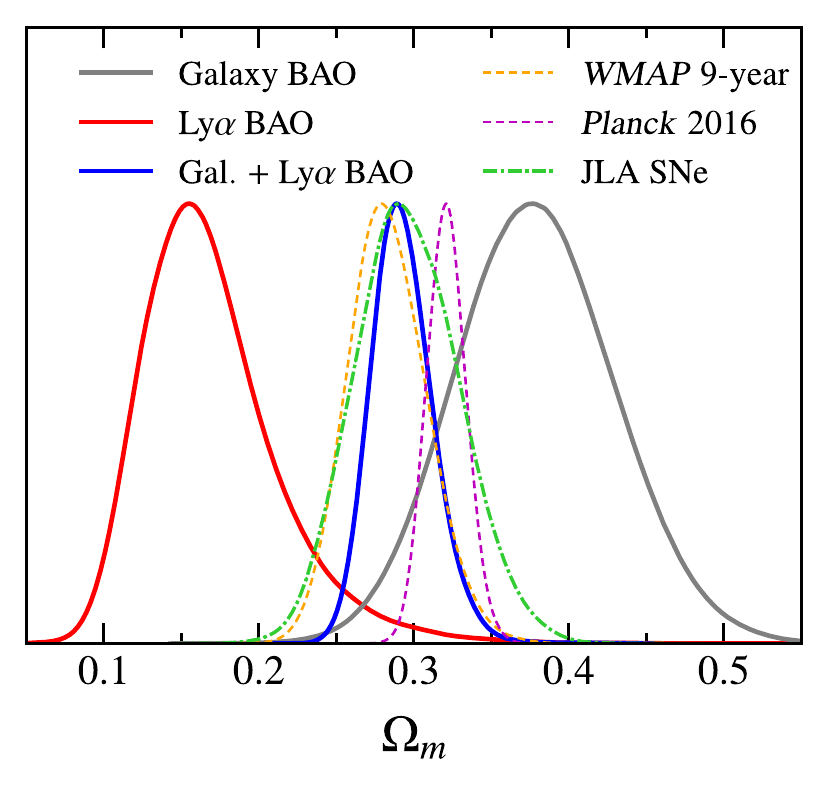}
\caption{{\em Left:} Comparison of BAO-only constraints in the flat \lcdm\ model. Contours containing 68 and 95\% of MCMC samples are shown for galaxy ($z_{\rm eff}\leq0.61$) and Ly$\alpha$ forest ($z_{\rm eff}\geq2.3$) BAO separately and in a joint fit using the BAO data listed in Table~1. In flat \lcdm\ the late-time expansion rate is determined only by $\Omega_m$, with $H_0r_d$ acting as an overall expansion normalization. {\em Right:} Comparison of $\Omega_m$ constraints from BAO, CMB and SNe measurements. The SNe constraint is from the ``joint light-curve analysis'' (JLA) presented by \cite{betoule/etal:2014}. While the combined BAO fit produces a tight constraint $\Omega_m=0.293\pm0.020$, in agreement with the CMB and SNe determinations, there is a $2.4\sigma$ tension between the galaxy and Ly$\alpha$ BAO, which individually prefer higher and lower values of $\Omega_m$, respectively.}
\end{figure*}

\subsection{Constraining $H_0$ with BAO plus deuterium abundance in \lcdm}

Obtaining a constraint on $H_0$ from the BAO requires adding information to break the $H_0-r_d$ degeneracy. One way to do this is to add a constraint on the baryon density \citep[e.g.,][]{addison/etal:2013b,aubourg/etal:2015,wang/xu/zhau:prep}. We assume that the photon energy density, or, equivalently, the CMB mean temperature, is also known. The CMB temperature was measured precisely by \COBE/FIRAS \citep{fixsen/etal:1996,fixsen:2009} and we view this result as independent of the CMB anisotropy measurements performed by more recent experiments. Note that while the $H_0$ in the $H(z)$ in the denominator of (1) cancels in the $H_0r_d$ product, some residual $H_0$ dependence still exists because the decoupling redshift and the sound speed depend on the {\em physical} matter and radiation densities, $\Omega_mh^2$ and $\Omega_rh^2$, respectively, while the expansion rate $H(z)$ depends on the {\em fractional} densities $\Omega_m$ and $\Omega_r$.

In the BAO fit with an external baryon density prior, $\Omega_m$ performs double duty. It not only goes into determining the late-time expansion ($D_M$ and $H$ at the BAO survey redshifts) but also controls the expansion history in the early universe prior to baryons decoupling from photons, since the photon and neutrino properties (with $N_{\rm eff}=3.046$) are held fixed. As well as providing an indirect $H_0$ constraint, the BAO+$\Omega_bh^2$ fit also serves as something of a self-consistency test of early and late-time expansion.

\begin{table*}
  \centering
  \caption{\lcdm\ constraints from the BAO+D/H fits, using either the theoretical or empirical $d(p,\gamma)^3\textrm{He}$ reaction rate, with CMB anisotropy constraints from \wmap\ and \planck\ included for comparison}
  \begin{tabular}{lcccc}
\hline
Parameter&BAO+D/H&BAO+D/H&\wmap\ 9-year&\planck\ 2016\\
&(theoretical)&(empirical)\\
\hline
$100\Omega_bh^2$&$2.156\pm0.020$&$2.257\pm0.034$&$2.265\pm0.049$&$2.215\pm0.021$\\
$100\Omega_ch^2$&$10.94\pm1.20$&$11.19\pm1.29$&$11.37\pm0.46$&$12.07\pm0.21$\\
$100\theta_{\rm MC}$&$1.0292\pm0.0168$&$1.0320\pm0.0173$&$1.04025\pm0.00223$&$1.04076\pm0.00047$\\
$H_0$~[km~s$^{-1}$~Mpc$^{-1}$]&$66.98\pm1.18$&$67.81\pm1.25$&$69.68\pm2.17$&$66.89\pm0.90$\\
$\Omega_m$&$0.293\pm0.019$&$0.293\pm0.020$&$0.283\pm0.026$&$0.321\pm0.013$\\
$r_d$ [Mpc]&$151.6\pm3.4$&$149.2\pm3.6$&$148.49\pm1.23$&$147.16\pm0.48$\\
\hline

\hline
\end{tabular}
\end{table*}

The most precise constraints on $\Omega_bh^2$ independent of the CMB power spectrum come from estimates of the primordial deuterium abundance. In standard Big Bang nucleosynthesis (BBN), the abundance of light nuclei is determined by a single parameter, the baryon-to-photon ratio $\eta$ \citep[see recent review by][and references therein]{cyburt/etal:2016}. Taking the photon number density as fixed from the CMB temperature, a measurement of the primordial deuterium abundance in conjunction with knowledge of BBN physics provides a constraint on $\Omega_bh^2$. Precise estimates of the primordial deuterium abundance have been made in recent years using extremely metal-poor damped Lyman-$\alpha$ (DLA) systems along sight lines to high-redshift quasars \citep[e.g.,][]{pettini/cooke:2012,cooke/etal:2014,cooke/etal:2016,riemer-sorensen/etal:2017}. \citeauthor{cooke/etal:2016} (2016; hereafter C16) report
\be
10^5\textrm{D}_{\rm I}/\textrm{H}_{\rm I}=2.547\pm0.033
\ee
by combining six such systems. The $d(p,\gamma)^3{\rm He}$ reaction rate plays a key role in the conversion from D/H to $\Omega_bh^2$. Using the theoretical calculation for this rate from \cite{marcucci/etal:2016}, C16 find
\be
\begin{split}
100\Omega_bh^2=2.156\pm0.020\\
(\textrm{D/H, theoretical rate}),
\end{split}
\ee
which is $>2\sigma$ lower than the \planck\ value (assuming a standard \lcdm\ model throughout). Using instead an empirically derived $d(p,\gamma)^3{\rm He}$ rate, C16 find
\be
\begin{split}
100\Omega_bh^2=2.260\pm0.034\\
(\textrm{D/H, empirical rate}),
\end{split}
\ee
which has a larger uncertainty but is in better agreement with CMB-derived values. We performed fits to the galaxy plus Ly$\alpha$ BAO data with the addition of each of the Gaussian priors on $\Omega_bh^2$ in (6) and (7) in turn. We show parameter constraints in Table~3, including the \wmap\ 9-year and \planck\ 2016 CMB anisotropy constraints for comparison.

\begin{figure}
\hspace{-0.3cm}
\includegraphics{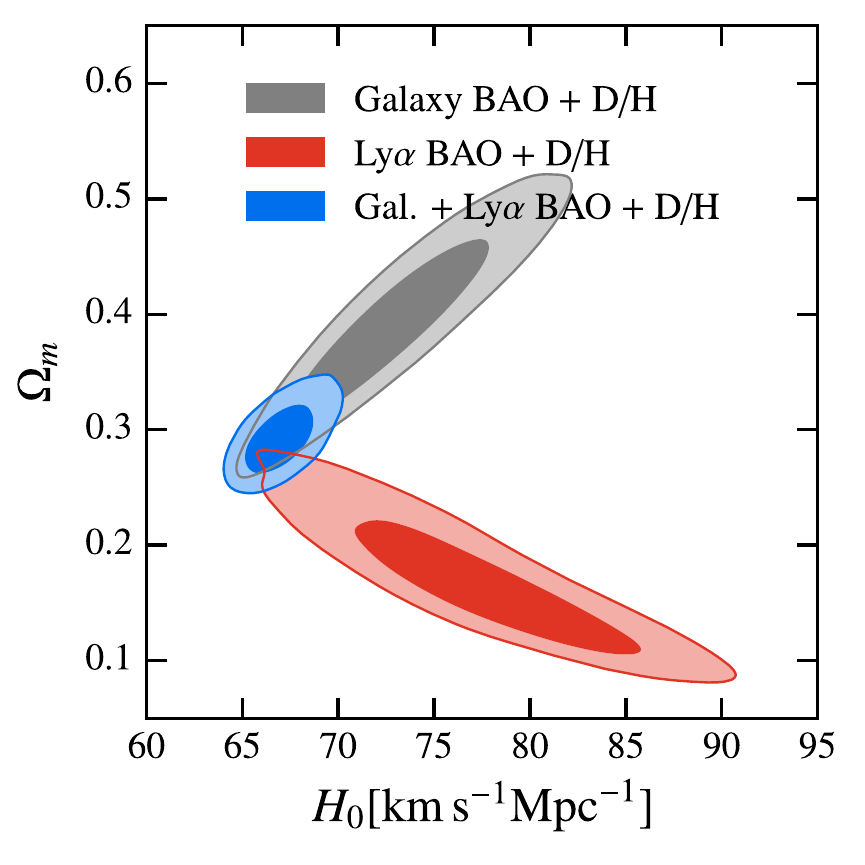}
\caption{Adding an estimate of the baryon density, $\Omega_bh^2$, in this case from deuterium abundance (D/H) measurements, breaks the BAO $H_0-r_d$ degeneracy in \lcdm. The same contours are shown as in Figure~3, with the addition of a Gaussian prior $100\Omega_bh^2=2.156\pm0.020$ \citep{cooke/etal:2016}. In contrast to Figure~3, here $\Omega_m$ determines both the early time expansion, including the absolute sound horizon, $r_d$, as well as the late-time expansion history. The radiation density is fixed from \COBE/FIRAS CMB mean temperature measurements. The combined BAO+D/H constraint, $H_0=66.98\pm1.18$~km~s$^{-1}$~Mpc$^{-1}$ is $3.0\sigma$ lower than the \cite{riess/etal:2016} distance ladder determination and is independent of CMB anisotropy data.}
\end{figure}

In the BAO+D/H fits, $\Omega_bh^2$ is driven solely by the D/H prior, as expected, and $\Omega_m$ matches the BAO-only value. While the choice of the $d(p,\gamma)^3{\rm He}$ reaction rate significantly impacts the value of $\Omega_bh^2$, it has a reduced impact on the inferred $H_0$, because $r_d$ only depends weakly on $\Omega_bh^2$ \citep{eisenstein/hu:1998,addison/etal:2013b}. Specifically, replacing the theoretical rate with the empirical one shifts the center of the $\Omega_bh^2$ distribution by 5.2 times the original uncertainty, but shifts the $H_0$ distribution by only 0.7 times the original uncertainty. Our BAO+D/H results for $H_0$ are more robust to the choice of rate than one might expect from the $\Omega_bh^2$ difference.

The $H_0$ values listed in Table~3 from the BAO+D/H fits have uncertainties of around 1.8\% and are $3.0$ and $2.5\sigma$ lower than the R16 distance ladder value of $73.24\pm1.74$~km~s$^{-1}$~Mpc$^{-1}$ for the theoretical and empirical $d(p,\gamma)^3{\rm He}$ rates, respectively. The combination of precise BAO and D/H measurements enables determinations of $H_0$ within the context of the flat \lcdm\ model that are almost 50\% tighter than the distance ladder measurement, and lower at moderate to strong significance. We emphasize that these constraints are completely independent of CMB anisotropy measurements.

Constraints in the $\Omega_m-H_0$ plane for the BAO+D/H fits with the theoretical $d(p,\gamma)^3{\rm He}$ rate are shown in Figure~4. We show results from the galaxy and Ly$\alpha$ BAO separately and together, as before. Tension between the galaxy and Ly$\alpha$ BAO is again apparent. Adding D/H to these data separately favors higher values of $H_0$, and it is only when galaxy and Ly$\alpha$ BAO are combined that $H_0$ is constrained to the values quoted in Table~3.

The direction of the Ly$\alpha$ BAO contour is roughly the same in the left panel of Figure~3 and in Figure~4, while that of the galaxy BAO contour changes. This behavior can be understood by considering how $r_d$ depends on $\Omega_m$ and $H_0$. For a given value of $\Omega_bh^2$, $r_d$ depends approximately on the combination $H_0\cdot\Omega_m^{1/2}$ \citep[equation 26 of][]{eisenstein/hu:1998}. This is the same dependence as $H(z)$ at the Ly$\alpha$ redshifts (Section~3.3) and is related to the fact that the universe is largely matter dominated in both cases. The dependence of $H(z)$ on $\Omega_m$ at the galaxy BAO redshifts is weaker, and the direction of the galaxy BAO contour in Figure~4 is approximately determined by requiring $H_0r_d$ to be roughly constant as $\Omega_m$ varies. This produces a positive correlation between $H_0$ and $\Omega_m$ because $r_d$ decreases as $H_0\Omega_m^{1/2}$ increases.

For the BAO+D/H fits, we ran \cosmomc\ as one would when fitting to the CMB: the fitted parameters are $\Omega_bh^2$, the physical cold dark matter density, $\Omega_ch^2$, and the angular sound horizon, $\theta_{\rm MC}$, and $H_0$, $\Omega_m$, and $r_d$ are derived from these three. Since the BAO+D/H data are insensitive to the amplitude and tilt of the primordial power spectrum, and the optical depth to reionization, these other \lcdm\ parameters are held fixed. Consistent results were obtained using earlier BAO and D/H data by \cite{addison/etal:2013b} and \cite{aubourg/etal:2015}. We note that \cite{riemer-sorensen/semjenssen:2017} recently obtained a tighter constraint on D/H than we have used here by combining the DLAs used by C16 with a number of additional measurements. Using this tighter constraint would not impact our conclusions.

\subsection{BAO and light element abundance constraints with varying $N_{\rm eff}$}

In the \lcdm+$N_{\rm eff}$ model, there is a perfect degeneracy between $\Omega_bh^2$ and $N_{\rm eff}$ from D/H measurements (Fig.~6 of C16). Closed contours in the $\Omega_bh^2-N_{\rm eff}$ plane can be obtained from combining estimates of the primordial D/H and $^4$He abundance \citep[e.g., review by][and references therein]{cyburt/etal:2016}. The primordial $^4$He abundance is estimated from He and H emission lines in extragalactic HII regions. Obtaining accurate constraints is challenging due to dependence on environmental parameters such as temperature, electron density, and metallicity, which must be modeled. An important recent development is the use of the HeI line at 10830~{\AA} to help break modeling degeneracies \citep{izotov/etal:2014}. The value of the primordial helium fraction reported by \cite{izotov/etal:2014}, $Y_p=0.2551\pm0.0022$, is, however, significantly higher than values found in some subsequent analyses of the same HII sample using different selection criteria and fitting methodology. For example, \cite{aver/olive/skillman:2015} found $Y_p=0.2449\pm0.0040$, while \cite{peimbert/peimbert/luridiana:2016} found $Y_p=0.2446\pm0.0029$. The different $Y_p$ values lead to significantly different inferences for $N_{\rm eff}$ when used in combination with D/H or CMB power spectra measurements. \cite{izotov/etal:2014} found evidence for additional neutrino species at 99\% confidence, while, for instance, \cite{cyburt/etal:2016} report $N_{\rm eff}=2.85\pm0.28$, and \cite{peimbert/peimbert/luridiana:2016} found $N_{\rm eff}=2.90\pm0.22$, consistent with the standard model value of 3.046.

Current D/H and $^4$He constraints clearly have the {\em precision} to weigh in significantly on the question of whether allowing $N_{\rm eff}>3$ is effective at resolving \lcdm\ tensions. Given the spread in $Y_p$ values discussed above, and the impact of the choice of $d(p,\gamma)^3{\rm He}$ rate when $N_{\rm eff}$ is allowed to vary (Section~5.2 of C16), we do not present a full set of results including BAO and light element abundance data for \lcdm+$N_{\rm eff}$. Instead we note that combining BAO measurements with D/H and $^4$He constraints on $N_{\rm eff}$ that are consistent with the standard model value would produce $H_0$ values consistent with the values in Table~3, although with larger uncertainties, while higher values of $N_{\rm eff}$ would produce a higher $H_0$, improving agreement with the distance ladder. The BAO measurements, being only sensitive to $H_0r_d$, and not to $H_0$ or $N_{\rm eff}$ directly, are unable to discriminate between these possibilities.

\section{Discussion}

We have presented evidence for a lower $H_0$ value than measured by the local distance ladder that is independent of \planck, both from combining BAO with other CMB datasets (\wmap, ACTPol and SPT), and from joint fits to BAO and D/H measurements, within the context of the standard \lcdm\ model. In light of this analysis it is clear that the $H_0$ tension cannot be resolved solely through a systematic error specific to the \planck\ data. It should be noted, however, that it is not simply a case of having a `high' $H_0$ from the distance ladder, and a `low' $H_0$ from \planck\ and the joint BAO fits. The high-multipole \planck\ temperature data prefer $H_0$ values that are even lower than the CMB+BAO or BAO+D/H values (bottom two rows of Table~2 and top right panel of Fig.~1). Restricting the \planck\ temperature power spectrum to multipoles $\ell>800$ produces
\be
\begin{split}
&H_0=65.12\pm1.45~\textrm{km}~\textrm{s}^{-1}~\textrm{Mpc}^{-1}\\
&(\textrm{\planck\ 2015 TT }\ell>800\textrm{, }\tau=0.07\pm0.02),
\end{split}
\ee
or
\be
\begin{split}
&H_0=64.30\pm1.31~\textrm{km}~\textrm{s}^{-1}~\textrm{Mpc}^{-1}\\
&(\textrm{\planck\ 2015 TT }\ell>800\textrm{, }\tau=0.055\pm0.009),
\end{split}
\ee
depending on the choice of $\tau$ prior. These values are not only in strong tension with R16, but are in moderate tension with some of the CMB+BAO values reported in Table~2. For example, for $\tau=0.055\pm0.009$, the SPT+BAO value is lower than R16 by $2.5\sigma$, but the \planck\ $\ell>800$ value is $2.6\sigma$ lower again than SPT+BAO. The shift in $H_0$ from adding the BAO to the $\ell>800$ \planck\ constraints is also larger than expected given the improvement in precision, as discussed in Section~3.1. Some $H_0$ tension remains even if we do not consider the distance ladder constraints. In fact, concordance cannot be achieved through the removal of {\em any} single data set (e.g., BAO, CMB, distance ladder, or D/H). This is part of the reason the $H_0$ discrepancy is challenging to resolve: a convincing solution must simultaneously address multiple avenues of disagreement.

A wide range of fits to expanded cosmological models, with various combinations of data, have been presented in recent years to try to reconcile $H_0$ and other parameter tensions. Our fits in this paper have been restricted to the standard flat \lcdm\ model, partly because our results for expanded models would be similar to those already presented by \cite{planck/13:2015}, \cite{alam/etal:2017}, \cite{heavens/etal:prep}, and others. The BAO, CMB, and light element abundance measurements have some common dependence on the early universe expansion history, which makes allowing freedom in, for example, $N_{\rm eff}$, seem attractive. As discussed in Section~1, varying $N_{\rm eff}$ does not sufficiently relieve tensions and is not statistically favored over standard \lcdm\ for the current BAO, CMB, and distance ladder data. There are good prospects for tightening $N_{\rm eff}$ constraints through improved measurements of the high-$\ell$ CMB damping tail in E-mode polarization \citep[e.g.,][]{snowmass/neutrino:2013,stage4:2016}. Future BAO data, for example from the Dark Energy Spectroscopic Instrument (DESI\footnote{http://desi.lbl.gov/}), Euclid\footnote{https://www.euclid-ec.org/}, and WFIRST\footnote{https://wfirst.gsfc.nasa.gov/}, will also provide significant improvements in precision over current measurements \citep[for BAO+$\Omega_bh^2$ forecasts for $H_0$, see][]{wang/xu/zhau:prep}.

\section{Conclusions}

We have examined the role of BAO measurements in determining $H_0$. While the BAO data alone are unable to distinguish between a change in $H_0$ and a change in the absolute sound horizon at decoupling, $r_d$, this degeneracy is broken, and a precise $H_0$ value obtained, when BAO are combined with either CMB power spectra or deuterium abundance measurements. Overall we find convincing evidence for a lower $H_0$ in \lcdm\ than obtained from the latest local distance ladder measurement \citep[$73.24\pm1.74$~km~s$^{-1}$~Mpc$^{-1}$;][]{riess/etal:2016} even without using data from \planck. The motivation and results of this study are summarized as follows:
\begin{enumerate}[label=(\roman*)]
\item Tension at the $>3\sigma$ level exists between determinations of $H_0$ from the distance ladder and the CMB anisotropy measurements from \planck, within the context of the standard flat \lcdm\ model. Other tensions also exist, for example between \planck\ data and constraints on the growth of structure from some weak lensing surveys.
\item None of the cosmological modifications commonly proposed appear to provide a statistically compelling solution to these tensions, although some, such as allowing freedom in the number of effective relativistic species, $N_{\rm eff}$, do reduce the $H_0$ disagreement.
\item Combining BAO measurements with CMB power spectrum data from \wmap, ACTPol, or SPT, produces $H_0$ values lower than the distance ladder by $2.4-3.1\sigma$, independent of \planck\ (Table~2). The difference was less pronounced in some earlier analyses because of using the angle-averaged BOSS CMASS BAO measurement. The angle-averaged $D_V(z)/r_d$ constraint is a compression of information and allows a region of parameter space that is ruled out by the full anisotropic BAO constraints (Fig.~2). Adding the BAO improves the $H_0$ constraint from ACTPol or SPT by more than a factor of three, making their precision comparable to the \planck-only results.
\item Combining BAO data with primordial deuterium (D/H) abundance estimates from metal-poor DLA systems produces precise $H_0$ values lower than the distance ladder by $2.5-3.0\sigma$, depending on assumptions about the $d(p,\gamma)^3{\rm He}$ reaction rate (e.g., $66.98\pm1.18$~km~s$^{-1}$~Mpc$^{-1}$ for the theoretical rate, see Table~3). This result is independent of any CMB anisotropy measurement and relies only on the CMB mean temperature from \COBE/FIRAS.
\item The two previous results taken together indicate that it is not possible to resolve the $H_0$ disagreement solely through some systematic error specific to the \planck\ data set.
\item The \planck\ high-multipole ($\ell>800$) damping tail measurements prefer lower values of $H_0$ than the combined BAO fits, for example $65.12\pm1.45$ and $64.30\pm1.31$~km~s$^{-1}$~Mpc$^{-1}$, for $\tau=0.07\pm0.02$ and $\tau=0.055\pm0.009$, respectively. The shift in $H_0$ from adding the BAO to these data is larger than expected at the $2.2$ and $2.8\sigma$ level for these $\tau$ priors. The $H_0$ disagreement is not as simple as the distance ladder value being `high' and other constraints coming out `low', and cannot be resolved through the removal of any single data set.
\item We note that a $2.4\sigma$ tension exists between the galaxy ($z\leq0.61$) and Ly$\alpha$ ($z\geq2.4$) BAO, as previously discussed by \cite{aubourg/etal:2015}. The BAO+D/H constraints rely on combining these measurements and as such it is important to review their consistency with future data.
\end{enumerate}

In recent years new precise measurements have led to multiple tensions, particularly in $H_0$, that are uncomfortably large to be explained by statistical scatter within the context of the standard \lcdm\ model. Whether this is the sign of new physics or underestimated uncertainties, or some combination of effects, remains unclear, and no straightforward explanation has yet presented itself. Near-term improvements in CMB, LSS, and distance ladder data are expected, however continuing to scrutinize existing measurements, as we have in this work, could also prove important in moving towards an eventual resolution.\\

\noindent
This research was supported in part by NASA grants NNX15AJ57G, NNX16AF28G, and NNX17AF34G, and JPL grant 1563692, and by the Canadian Institute for Advanced Research (CIFAR). We acknowledge the use of the Legacy Archive for Microwave Background Data Analysis (LAMBDA), part of the High Energy Astrophysics Science Archive Center (HEASARC). HEASARC/LAMBDA is a service of the Astrophysics Science Division at the NASA Goddard Space Flight Center. This work was partly based on observations obtained with \planck\ (\url{http://www.esa.int/Planck}), an ESA science mission with instruments and contributions directly funded by ESA Member States, NASA, and Canada. Part of this research project was conducted using computational resources at the Maryland Advanced Research Computing Center (MARCC).

The authors would like to thank Erminia Calabrese, Antony Lewis, James Rich, Adam Riess, and Ashley Ross for helpful discussions and clarifications regarding data sets and software used in this work, and we are also grateful to Adam Riess for reading a draft of the manuscript and providing useful suggestions. We acknowledge the use of the \texttt{GetDist} plotting package\footnote{http://getdist.readthedocs.io/en/latest/}.
\\
\\
\bibliographystyle{apj}

\end{document}